\documentclass{article}

\usepackage{arxiv}

\usepackage[utf8]{inputenc} 
\usepackage[T1]{fontenc}    
\usepackage{hyperref}       
\usepackage{url}            
\usepackage{booktabs}       
\usepackage{amsfonts}       
\usepackage{nicefrac}       
\usepackage{microtype}      
\usepackage{lipsum}
\usepackage{graphicx}
\graphicspath{ {./images/} }

\title{Agentic TinyML for Intent-aware Handover in 6G Wireless Networks}

\author{
 Alaa Saleh \\
  Center for Ubiquitous Computing\\
  University of Oulu\\
  Oulu 90014, Finland \\
  \texttt{alaa.saleh@oulu.fi} \\
   \And
 Roberto Morabito \\
  Department of Communication Systems\\
  EURECOM\\
  Biot 06410, France \\
  \texttt{roberto.morabito@eurecom.fr} \\
  \And
 Sasu Tarkoma \\
  Department of Computer Science\\
  University of Helsinki\\
  Helsinki 00100, Finland \\
  \texttt{sasu.tarkoma@helsinki.fi} \\
  \And
 Anders Lindgren \\
  RISE Research Institutes of Sweden\\
  Stockholm 166 40, Sweden \\
  Department of Computer Science, \\
  Electrical and Space Engineering\\
  Luleå University of Technology \\
  Luleå 971 87, Sweden\\
  \texttt{anders.lindgren@ri.se} \\
    \And
 Susanna Pirttikangas \\
  Center for Ubiquitous Computing\\
  University of Oulu\\
  Oulu 90014, Finland \\
  \texttt{susanna.pirttikangas@oulu.fi} \\
    \And
 {L}auri Lov\'en \\
  Center for Ubiquitous Computing\\
  University of Oulu\\
  Oulu 90014, Finland \\
  \texttt{lauri.loven@oulu.fi} \\
}

\begin{document}
\maketitle
\thispagestyle{fancy}
\fancyhf{}  
\lfoot{\scshape Preprint. Under review.} 
\cfoot{\thepage}  

\pagestyle{empty}

\begin{abstract}
As 6G networks evolve into increasingly AI-driven, user-centric ecosystems, traditional reactive handover mechanisms demonstrate limitations, especially in mobile edge computing and autonomous agent-based service scenarios. This manuscript introduces WAAN, a cross-layer framework that enables intent-aware and proactive handovers by embedding lightweight TinyML agents as autonomous, negotiation-capable entities across heterogeneous edge nodes that contribute to intent propagation and network adaptation. To ensure continuity across mobility-induced disruptions, WAAN incorporates semi-stable rendezvous points that serve as coordination anchors for context transfer and state preservation. The framework’s operational capabilities are demonstrated through a multimodal environmental control case study, highlighting its effectiveness in maintaining user experience under mobility. Finally, the article discusses key challenges and future opportunities associated with the deployment and evolution of WAAN.
\end{abstract}


\section{Introduction}
We are witnessing the rapid emergence of user-centric, edge-based ecosystems where 6G wireless networks are expected to host an increasing number of agentic services~\cite{brinton2025key}. These services, powered by autonomous AI agents, introduce rising computational and communication requirements that must be met under highly dynamic and heterogeneous environments. As a result, even fundamental mechanisms such as routing, task scheduling, and offloading are evolving beyond traditional models, moving toward more adaptive, context-aware strategies capable of supporting distributed decision-making across mobile devices and edge servers~\cite{zhou2024deploying}.

This evolution is also driven by the fact that future 6G networks will be increasingly AI-driven and intent-driven~\cite{brinton2025key}, with decision-making and service orchestration shifting closer to the edge. As a result, traditional network functions, including the concept of handover itself, need to be revisited. In addition to classical handovers that only transfer connectivity, emerging agentic ecosystems call for a new form of \textit{intent handover} where the focus is on seamlessly transferring the execution context of user intents across AI agents rather than simply maintaining a radio link. In this respect, classic reactive handovers are no longer sufficient to cope with the demands of highly dynamic environments and resource-constrained edge nodes~\cite{saleh2023towards}. Moreover, the combination of user mobility, traffic fluctuations, and heterogeneous device capabilities introduces additional layers of complexity~\cite{jiang2024large}, making resilience (i.e., the ability to maintain or restore service continuity under disruption) a critical concern in these ecosystems. When users move, intermittent connectivity can easily lead to delayed or lost responses~\cite{baldoni2009distributed}. These challenges highlight the need to rethink how 6G architectures and AI agents work together to deliver seamless, intent-driven services in mobile environments.

Building on this vision, in such dynamic wireless edge environments, the architectural and operational foundations of 6G systems will directly shape how AI agents operate in mobile and distributed settings. These foundations will aim to deliver user-centric services that seamlessly adapt to changes in connectivity, location, user preferences, and network conditions, ensuring responsive and personalized experiences~\cite{zeng2024implementation}. Achieving this adaptability requires more than static mechanisms: it depends on effective inter-agent negotiation across edge servers and mobile devices~\cite{xue2025wdmoe,xu2024large}. In addition, leveraging quality-of-experience (QoE) metrics and user feedback~\cite{lee2025ai} becomes essential to refine decision-making, responsiveness, and resilience.

\begin{figure}[!t]
  \centering
  \includegraphics[width=0.6\linewidth]{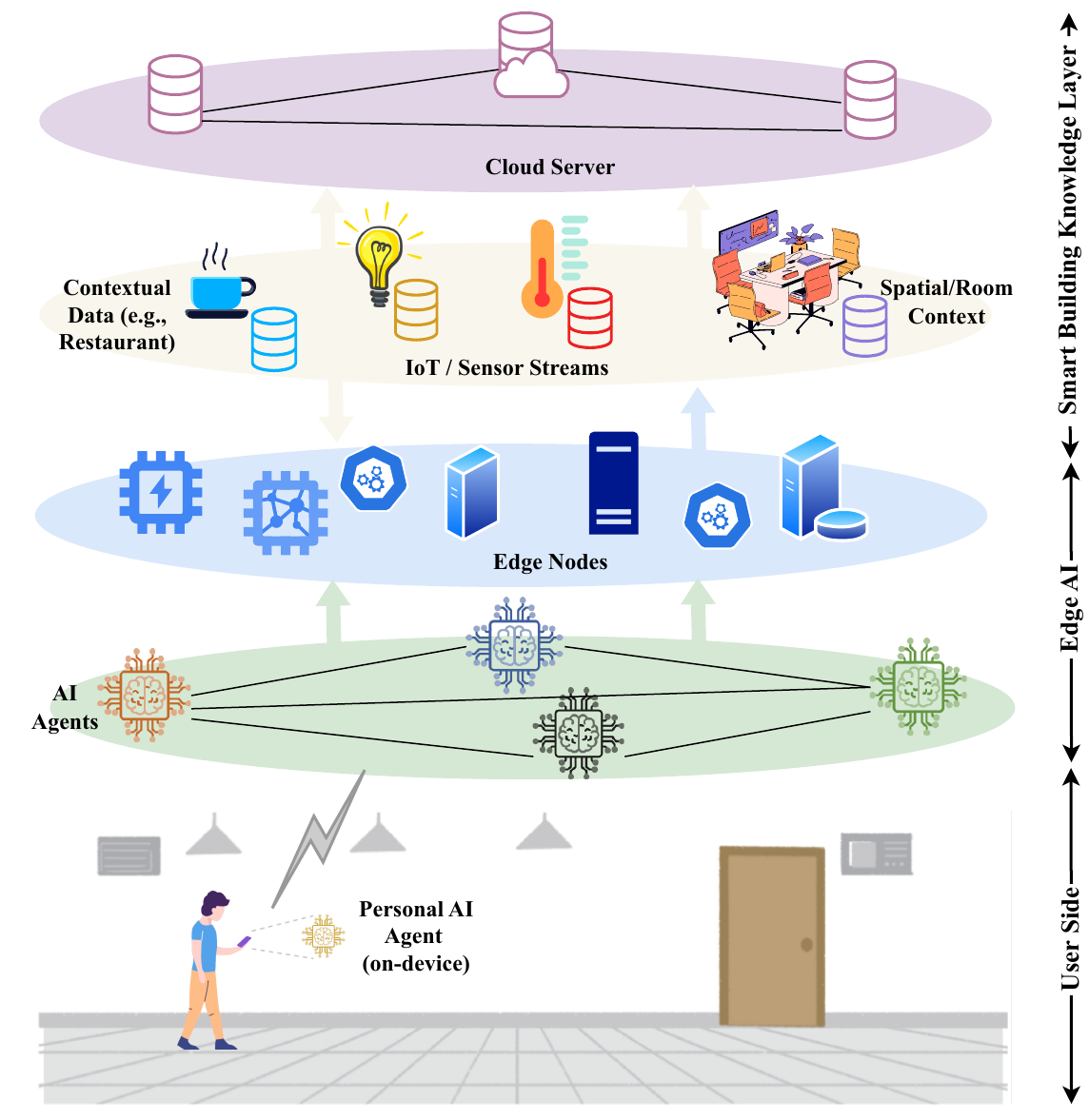}
  \caption{Illustration of the intent-driven processing workflow in heterogeneous wireless environments. A user submits an intent that is decomposed by a personal agent on the user’s device and offloaded to a nearby AI agent running on an edge device. The AI agent retrieves data from smart building datasets (e.g., rooms, restaurant, local sensors) to process the intent and returns the response to the user. User mobility can disrupt this process if handover is not supported.}
  \label{fig:waan}
\end{figure}

To illustrate these challenges, Fig.\ref{fig:waan} shows a typical workflow: when a user submits an intent, a personal agent on the user's device decomposes the request into sub‑tasks and forwards them to the closest AI agent running on a nearby edge device. This AI agent processes the sub‑tasks by retrieving data from relevant sources (e.g., smart building datasets containing information from rooms, restaurants, or local sensors) and generates a response, which is then returned to the user as a notification or action. However, user mobility can interrupt this workflow: if the user moves out of range during processing, the connection between the personal agent and the AI agent is lost. Without a mechanism for seamless continuation, the request must be resubmitted to a new AI agent, forcing redundant recomputation and causing additional latency, power consumption, and overhead.

This scenario highlights a key limitation: current systems lack mechanisms to seamlessly transfer the execution context of user intents across AI agents as users move. Overcoming this limitation requires rethinking how mobility is handled in AI-driven, intent-driven 6G networks, a challenge that we address in the next section by introducing the Wireless AI Agent Network (WAAN).

\section{Introducing the Wireless AI Agent Network (WAAN)}
In the WAAN, we envision an AI interconnect layer of agents operating across heterogeneous nodes to autonomously and seamlessly support the continuation of user intents across dynamic edge environments. Rather than focusing solely on connectivity, WAAN agents coordinate and negotiate task execution to maintain the semantic and computational continuity of user intents as they move. Crucially, this propagation process is not confined to the application or service layer—it operates in a cross-layer fashion, adapting to real-time wireless conditions, device capabilities, and network-level metrics~\cite{tarkoma2010overlay,saleh2025follow}. For instance, local AI models (e.g., Small Language Models (SLMs) or Tiny Machine Learning (TinyML) agents) can adjust task offloading, scheduling, or response generation based on variations in signal quality, congestion, or resource availability. Through this cross-layer adaptive intelligence, spanning both the device and network layers, WAAN aims to reduce redundant processing, minimize routing overhead, and support proactive intent handovers. While this represents a conceptual architecture, it addresses limitations in current systems that treat intent propagation as isolated from the underlying network dynamics.

For enabling this vision, WAAN is conceived as a loosely coupled, policy-driven architecture where agents collaborate without requiring tight synchronization, allowing the system to remain robust in dynamic, heterogeneous edge environments. A key design requirement is the deployment of adaptive intelligence on resource-constrained nodes, where lightweight AI and ML models continuously learn from local observations and neighbor interactions. This distributed intelligence allows agents to negotiate with one another, balancing their own intents and resource limitations with system-wide goals~\cite{10.24963/ijcai.2024/890}. 
Efficient negotiation requires continuously learning agent behaviors that can autonomously manage service interruptions, reroute around failures, optimize resource usage, and dynamically adapt to evolving user intents. These capabilities are essential for maintaining a high quality of user experience under conditions of fluctuating connectivity, mobility, and network congestion. Central to this process is the intelligent management of how data is transmitted over the wireless channel~\cite{chen2024big}. This clearly involves the effective scheduling of wireless resources and the allocation of time–frequency radio resources to traffic flows, but also the ability to adapt to the nature of the applications generating that traffic. For example, Generative Artificial Intelligence (GenAI)-powered agents are expected to drive a surge of uplink and downlink traffic due to interactive video assistants and immersive, multimodal applications, further straining network infrastructure. At the same time, the same GenAI models can help mitigate this load in closed-edge deployments by processing and semantically compressing data closer to the user. The possibility of extracting and transmitting only the most relevant content through these models can significantly reduce the volume of data sent over the network, alleviating congestion while maintaining responsiveness and quality~\cite{saleh2023towards}.

Compared to existing agentic frameworks, which primarily focus on semantic reasoning and coordination among agents at the application layer, WAAN explicitly integrates cross-layer awareness and proactive intent handover. In WAAN, intent handover is not limited to passing control between agents of the same type: it spans different agentic domains (inter-agent systems), computing tiers (from on-device to edge to cloud), and data and knowledge sources on which these agents operate (as illustrated in Fig. \ref{fig:waan}). This enables an ongoing task to continue seamlessly even as the user moves across heterogeneous environments, devices, and datasets. The idea is to link decision-making to network state, making WAAN agents able to adapt task routing, offloading, and intent propagation strategies based on real-time wireless conditions. This combination of semantic intent handling with network-level adaptability distinguishes WAAN from current agentic systems, which largely treat networking as a passive substrate.

\begin{figure}
  \centering
  \includegraphics[width=0.9\linewidth]{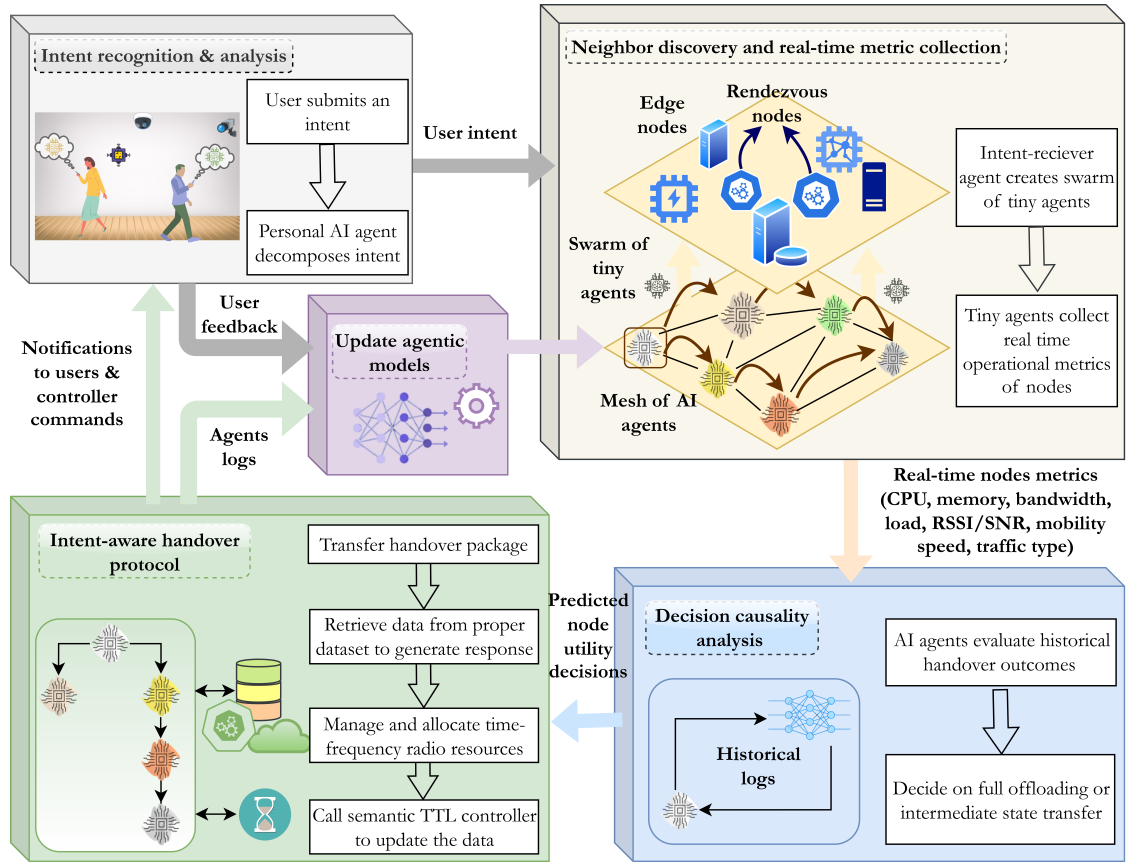}
  \caption{WAAN architecture for intent-aware handovers.
User intents are decomposed by personal agents and processed within a mesh of AI agents and swarms of tiny agents deployed on heterogeneous edge nodes. Tiny agents continuously collect real-time metrics (CPU, bandwidth, mobility, etc.) to guide routing, resource allocation, and intent handovers. The intent-aware handover protocol transfers execution context and intermediate results, while the decision causality module leverages historical logs to decide between full offloading and intermediate state transfer. Feedback loops update agentic models over time, ensuring adaptive, generalizable intelligence aligned with user QoE requirements.} 
  \label{fig:architecture}
\end{figure}

\section{The Role of TinyML in WAAN}
The realization of WAAN depends on the ability of even the most constrained devices to perform local inference and decision-making without relying on powerful cloud resources. TinyML offers lightweight, energy-efficient models that can run on a wide range of end and edge devices, from microcontrollers and low-power processors to more powerful AI-accelerated single-board computers, enabling agents to perceive context and act locally. In the WAAN framework, TinyML allows each node to autonomously evaluate conditions such as connectivity, load, and user intents, and to participate in negotiation processes without incurring prohibitive communication delays.

These models are not limited to application-layer reasoning. Instead, TinyML agents incorporate cross-layer signals, ranging from channel quality to energy availability, to inform decisions on when to offload tasks, when to handover intents, and how to manage local resources. Additionally, thanks to the possibility of relying on lightweight few-shot learning mechanisms, TinyML agents can adapt to new traffic patterns and environmental changes using minimal training data, thereby maintaining responsiveness even under previously unseen conditions \cite{rajapakse2023intelligence}.

A fundamental aspect of WAAN is its ability to adapt to the heterogeneous capabilities of participating devices. Extremely constrained nodes, such as microcontrollers, cannot run multiple or complex models simultaneously, which limits the type of decisions they can take locally. To cope with this limitation, WAAN introduces agent capability discovery mechanisms that allow agents to be aware of their own strengths and weaknesses and to collaborate with other agents accordingly \cite{11022747}. Through this awareness and cooperation, even devices with minimal capabilities can contribute meaningfully to compound functions, relying on more capable peers for tasks beyond their local capacity. Conversely, more powerful devices, such as smartphones and edge nodes, can host multiple models and take on more complex reasoning or coordination roles, while still benefiting from distributed collaboration.


Looking at the new GenAI wave, TinyML can well complement more powerful models in WAAN by providing fast, local reactions, while complex reasoning or semantic understanding—such as intent interpretation or multimodal data summarization—can be offloaded to more capable agents. This hierarchical collaboration, deployable on emerging edge AI platforms, ensures that only the most critical or resource-intensive tasks reach the edge or cloud, while real-time decisions, such as choosing the best next hop for an intent, remain at the constrained node.

Our vision places TinyML as an active participant in a distributed agentic ecosystem. This approach is rather different than typical TinyML-based systems, which usually act in isolation for sensing or classification tasks. Through WAAN, TinyML nodes evolve from passive edge sensors to autonomous, negotiation-capable agents that play a direct role in intent propagation and network adaptation. This integration can represent an important paradigm shift for realizing intent-driven, cross-layer services in future 6G networks.

\section{Generalizable and Adaptive Intelligence: An In-Depth Look at WAAN}
\vskip -1\baselineskip
The WAAN forms an AI interconnect layer composed of autonomous agents distributed across heterogeneous network nodes. This intelligent layer autonomously supports user mobility and enables context-aware intent handovers through three main operational strategies: \emph{(i)} bandwidth-aware resource allocation, \emph{(ii)} dynamic inference offloading, and \emph{(iii)} selective invocation of computational modules. Together, these strategies aim to optimize system performance under resource-constrained and dynamically changing conditions (see Fig. \ref{fig:architecture}).

Fig. \ref{fig:architecture} provides an overview of the WAAN architecture. At the top-left, user intents are recognized and decomposed by a personal AI agent. These intents are then processed through a mesh of AI agents (center and right), where tiny agents continuously perform neighbor discovery and real-time metric collection on parameters such as CPU load, memory usage, bandwidth availability, mobility patterns, and traffic type. The intent-receiver agent dynamically creates a \textit{swarm} of tiny agents that gather these metrics from heterogeneous edge nodes, enabling decisions that are both context-aware, knowledge-aware, and mobility-aware.

Edge nodes, however, are resource-constrained and often exposed to new user intents and operational patterns, such as sudden surges of immersive XR traffic, new types of GenAI assistants, or unexpected context changes caused by user mobility. Retraining large models on such nodes is infrequent due to overhead, which motivates the WAAN framework to rely on a self-expanding mesh of lightweight tiny-agents that collaboratively develop agentic policies guiding decision-making across the network. These agents need to track the operational requirements of each node but also predict the most suitable neighboring nodes for intent handover by analyzing the causal dependencies of past routing decisions. This capability is further enhanced by \textit{few-shot generalization}, where minimal examples of success or failure in similar contexts allow the agents to make confident predictions for new scenarios.

The idea of embedding few-shot learning capabilities into TinyML models stems from the need to try making the intelligence capabilities of WAAN as generalizable as possible and at scale. Agents can adapt their negotiation mechanisms and improve local decision-making processes even in highly dynamic traffic conditions. Within this architecture, agents operate as autonomous control entities, responsible for scheduling tasks, managing computational offloading, and balancing edge resource loads against the predicted utility of each action to ensure resource allocation that maximizes benefit.


To support these adaptive behaviors, WAAN integrates dynamic offloading strategies, allowing tasks to be either fully offloaded or partially transferred via intermediate computational states. This enables new nodes to resume execution from prior states rather than reinitiating the task from scratch, thereby reducing latency, overhead, and energy consumption during mobility-induced disruptions. However, effective execution of such stateful handovers requires architectural elements that can maintain semantic continuity and state preservation across mobility-induced disruptions during agentic and intent handovers.

To address this, rendezvous points in WAAN are envisioned to serve as semi-stable coordination nodes deployed at fixed edge or cloud locations. These nodes act as anchor points where user intents, data streams, and agentic policies intersect within the network. In our vision, these rendezvous points could become the places where contextual state and intent metadata are temporarily cached or synchronized, potentially reducing the risk of interruptions and avoiding unnecessary recomputation when a handover occurs. At the same time, they could also provide an opportunity to maintain a certain level of system accountability and auditability: by concentrating part of the coordination at specific points in the network, it becomes feasible to log decisions, data exchanges, and agent behaviors, which is much harder to achieve in a purely fully distributed setup. We see these rendezvous points as one possible way of introducing some form of stateful coordination in what would otherwise be a completely dynamic and volatile agentic environment.

Historical logs and experiential feedback loops (lower right of Fig. \ref{fig:architecture}) enable agents to evaluate handover outcomes, recover from failed attempts, and fine-tune their internal models over time, which leads to better predictive decisions for future routing and intent placement. Additionally, fallback candidates may be pre-ranked, allowing for rapid reassignment when primary handover fails.

Finally, intent-based handovers are managed through an \textit{intent propagation protocol} that incorporates a \textit{semantic time-to-live (TTL)}. Unlike traditional TTLs based purely on time, semantic TTL also reflects contextual relevance. This ensures that the temporal and contextual relevance of information is taken into account when transferring intents, so that decisions on routing and offloading are always aligned with user QoE requirements such as latency sensitivity and accuracy.

\section{Case Study: Intent Handover with Cross-Layer Adaptation}
We now illustrate how the WAAN architecture operates in practice, by considering representative case studies that highlight the role of intent handover, TinyML-driven cooperation, and cross-layer adaptivity. With these scenarios, we want to demonstrate how WAAN agents handle user mobility and heterogeneity while ensuring seamless continuation of tasks without recomputation or service degradation.

Consider a user carrying a 6G-enabled smartphone who submits a complex intent: \textit{"Provide a live multimodal summary of the room I am entering (using camera, microphone, and IoT sensors) and adapt the environment accordingly (lighting, temperature)"}. The user’s personal agent decomposes this intent into subtasks such as sensor fusion, multimodal summarization, and environment control, and offloads them to the closest WAAN agent based on current location. As the user moves from \textit{Zone A} to \textit{Zone B}, they traverse several WAAN coverage areas, each supported by edge nodes running autonomous AI agents that process these intents in real time.

\begin{figure}[!t]
  \centering
  \includegraphics[width=0.93\linewidth]{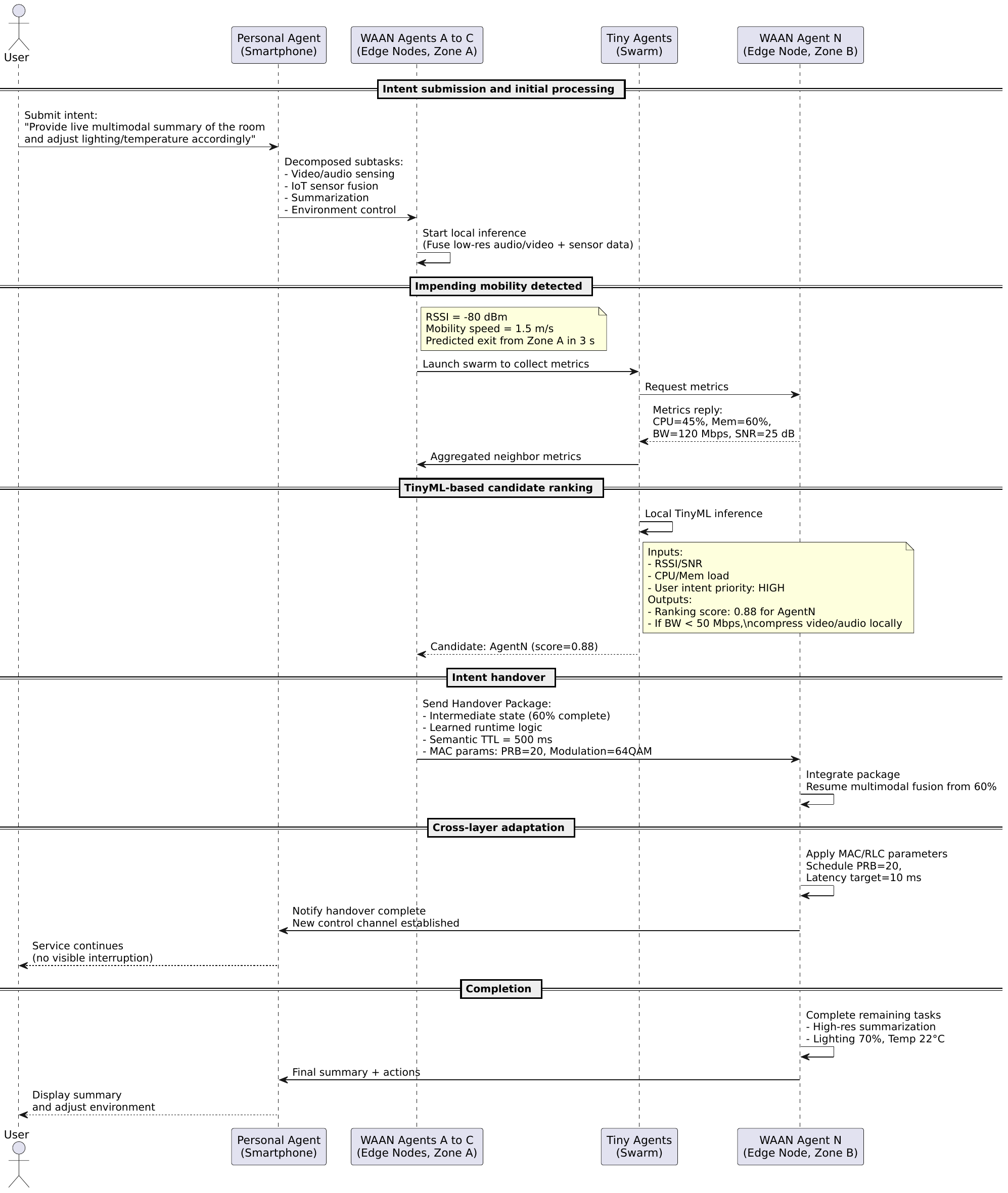}
  \caption{Sequence diagram showing intent-aware handover with cross-layer adaptation in WAAN. A multimodal summarization task is handed over between agents as the user moves, without restarting, guided by TinyML-based ranking and context transfer.}
  \label{fig:handover}
\end{figure}

During this mobility, the initial AI agent (Agent A) may detect that the user is about to leave its coverage. Rather than allowing the session to drop, Agent A triggers an intent-aware handover protocol (illustrated in Fig. \ref{fig:handover} and supported by the WAAN architecture in Fig. \ref{fig:architecture}). This protocol uses a swarm of TinyML-powered agents to collect real-time operational metrics—such as CPU load, memory usage, available bandwidth, RSSI/SNR, mobility speed, and current traffic type—from neighboring nodes. Using these metrics, the tiny agents rank potential target agents using lightweight, local inference models. This decision incorporates both application-level requirements (e.g., low latency for multimodal processing) and network-level conditions (e.g., congestion, link quality).

Based on this ranking, Agent A selects a new target (Agent N) and performs a knowledge-driven handover. Instead of simply transferring unfinished subtasks, Agent A sends a handover package that includes: \emph{(i)} the intermediate task state (e.g., 60\% of multimodal summarization already complete), \emph{(ii)} the refined runtime logic and learned policies, \emph{(iii)} the semantic TTL of context data, and \emph{(iv)} relevant MAC/RLC parameters. This package ensures that the receiving agent can continue execution exactly where it left off, without recomputation, while also inheriting insights from Agent A’s past decisions.
If Agent N becomes unreachable during transfer, the system can quickly fall back to the next-ranked candidate, preserving progress with minimal degradation.

Upon receiving this package, Agent N integrates the logic into its own runtime, applies the MAC/RLC parameters for optimized scheduling of time-frequency resources, and re-establishes a control channel with the user’s personal agent. This process exemplifies cross-layer adaptation: the application intent (multimodal environment awareness and control) and the lower-layer decisions (radio scheduling and resource allocation) are handled in a coordinated way, driven by TinyML at the lower tiers and larger models on edge nodes when needed.

After the handover, Agent N validates the relevance of the computed results using the semantic TTL, ensuring that stale data does not degrade quality. The response is then sent to the personal agent, which delivers a seamless user experience with no noticeable interruption. Through this integration of intent propagation, TinyML-driven agent cooperation, and cross-layer adaptation, WAAN enables a system that learns continuously and remains responsive under diverse mobility and environmental conditions.

This example is representative of the broader class of scenarios that WAAN targets: tasks where user mobility, heterogeneous resources, and dynamic network conditions can easily disrupt (complex) application-level experiences. It illustrates how WAAN’s combination of knowledge-driven handover, TinyML-powered cooperation, and cross-layer adaptivity can preserve computational state and context while ensuring seamless service continuity. Building on this representative scenario, the next sections discuss the advantages introduced by WAAN, the challenges that remain, and future research directions for intent-driven 6G edge networks.

\section{Advantages}
The WAAN’s explicit integration of semantic, intent-level processing with real-time wireless and network state enables agents to optimize routing, task offloading, and resource allocation. Through cross-layer adaptive propagation, WAAN reduces redundant computation, minimizes routing overhead, and supports proactive intent handovers that ensure computational continuity. During mobility events, proactive offloading and intermediate-state transfers, orchestrated through rendezvous points as semi-stable coordination anchors, allow receiving nodes to resume execution without recomputation, thereby reducing disruption-induced delays and bandwidth overhead while maintaining user QoE under dynamic network conditions.

Furthermore, agentic behaviors, driven by continuous learning from local observations and neighbor interactions, could enhance adaptive intelligence across resource-constrained nodes. This may support dynamic intent adaptation, efficient resource management, and enhanced resilience by enabling service continuity in the presence of node failures, mobility, or other disruptions.
Moreover, by incorporating cross‑layer signals and relying on few‑shot adaptation, TinyML may further extend autonomous control at the edge. Consequently, TinyML agents can intelligently decide when to offload, hand over intents, and allocate local resources efficiently, all while maintaining responsiveness and reducing reliance on distant compute infrastructures.

\section{Challenges and Future Research}
Despite the advantages of WAAN in adaptability, efficiency, and autonomy, realizing this vision raises a number of fundamental challenges. These challenges arise from the heterogeneity of devices and models, the dynamics of wireless environments, and the lack of mature mechanisms for large-scale agent coordination. Addressing them will require new ideas in learning, orchestration, and security to fully unlock the potential of agentic intelligence in 6G edge networks.
\\
\\
\textbf{Semantic transfer and intent handover.} Coordinating the transfer of execution context—including models, intermediate states, runtime logic, and learned policies—across heterogeneous agents is complex in mobile environments. While emerging agent platforms include basic task handoff primitives, they are designed for static conditions and do not yet support intent-driven orchestration under wireless constraints.
\\
\\
\textbf{Intent coordination and user-in-the-loop adaptation.} As WAANs scale, overlapping user intents will compete for shared resources and impose heterogeneous latency and compute demands. Large language models may act as brokers to interpret and prioritize intents, while continuous user feedback can refine agent behavior and align it with evolving preferences.
\\
\\
\textbf{Cross-layer reasoning.} Agents must learn to reason jointly across application, transport, and radio layers, adapting decisions to current network and resource conditions. Achieving this while balancing TinyML models on constrained nodes with larger models on more capable devices requires new coordination and resource allocation strategies.
\\
\\
\textbf{Learning under uncertainty.} Applying WAAN in unstructured or highly dynamic environments remains challenging. Few-shot learning methods must be extended to generalize under new mobility patterns, network behaviors, and traffic distributions, enabling rapid adaptation despite limited data.
\\
\\
\textbf{Protocols, security, and compliance.} WAANs require new standardized protocols that combine application-level semantics with network context, while also ensuring trust, security, and regulatory compliance. Rendezvous points may help provide accountability and safe handling of sensitive state during intent handovers, but additional mechanisms will be needed to satisfy privacy regulations such as GDPR and the AI Act.

\section{Conclusion}
This article introduced WAAN as a cross-layer adaptive intelligence architecture designed to enable intent-aware and proactive handovers, representing a shift toward generalizable, intent-driven services in 6G agentic systems. By integrating lightweight TinyML agents with cross‑layer decision mechanisms spanning device and network layers, WAAN achieves seamless continuity of user intents under mobility, heterogeneous resources, and fluctuating wireless conditions. The framework links decision‑making to real‑time network state, enabling agents to adapt task routing, offloading, and intent propagation while reducing redundant processing and routing overhead. Moreover, the integration of rendezvous points as semi-stable coordination nodes enhances continuity in WAAN by enabling continuity and state preservation, ensuring intent-aware orchestration across dynamic 6G environments. Finally, while WAAN architecture introduces a robust basis for 6G agentic services, future research is required to address challenges related to semantic state transfer, adaptive learning under distribution shifts, and standardization of agent integration protocols.

\section*{Acknowledgment}
This research is funded by the Research Council of Finland through the 6G Flagship (Grant Number 369116) project, and by Business Finland through the Neural Pub/Sub project (Diary No. 8754/31/2022) and the Digital Twinning of Personal Area Networks for Optimized Sensing and Communication project (Diary No. 8782/31/2022).

\bibliographystyle{unsrt}  
\bibliography{references}  

\end{document}